\documentclass[aps,11pt,preprintnumbers,nofootinbib,floatfix]{revtex4}

\usepackage{graphicx}
\usepackage{epsfig}
\usepackage{dcolumn}
\usepackage{subfigure}
\usepackage{multirow}
\usepackage{amsmath}
\usepackage{amssymb}

\begin{document}

\title{Effects of string cloud on Gauss-Bonnet holographic superconductors}


\author{Cao H. Nam}
\email{hncao@yonsei.ac.kr} 
\affiliation{Institute of Research and Development, Duy Tan 
University, Da Nang 550000, Vietnam}

\begin{abstract}%
The effects of the string cloud on higher-dimensional holographic superconductors in Einstein-Gauss-Bonnet gravity are investigated in the probe limit. The critical temperature is analytically obtained using Sturm-Liouville eigenvalue method. It is observed that the critical temperature only exists in an allowed region of the parameter space. Also, the presence of the string cloud with the sufficiently large density should prevent the the existence of the critical temperature. As increasing the string cloud density parameter, the critical temperature decreases in the region of the sufficiently low charge density but increases in the region of the sufficiently high charge density. Whereas, the presence of Gauss-Bonnet terms always makes the critical temperature decreasing. In addition, the expression of the condensation operator and the critical exponent are computed analytically.
\end{abstract}

\maketitle

\section{Introduction}
The AdS/CFT correspondence \cite{Maldacena}, which relates a weakly coupling gravity theory in AdS spacetime to a strongly coupling field theory on the boundary of AdS spacetime, has attracted consideration attention because of its computational power. According this correspondence, one can calculate the quantities of the strongly coupling field theory by using the gravitational dual in one higher dimension. In recent years, the AdS/CFT correspondence has been used to study various phenomena in condensed matter physics \cite{Hartnoll2010}. In particular, much attention have been dedicated to apply the holographic method for superconductors because high temperature superconductors are in the strong coupling regime and thus can not be described by the BCS theory. The dual gravitational description of $s$-wave superconductors was first proposed by Hartnoll et al. at which a phase transition from a black hole with no hair to one with scalar hair is interpreted as the (normal) conductor/superconducting phase transition in the dual field theory \cite{Horowitz2008a,Horowitz2008b}. Following this work, the investigations of holographic superconductors in the framework of Einstein gravity have attracted a lot of attention \cite{Rodriguez-Gomez2010,Jing2010,Therrien2010,Yang2010,Pan2011,Zong2011,Zhang2011,GeLeng2012,
Gangopadhyay2012,Jing2013,Erdmenger2013,Lai2015,Ghorai2016,Liu2016,Zangeneh2016,
Sheykhi2016,Salahi2016,Ghazanfari2018,Asl2018}.

At short distances, it is expected that Einstein gravity would obtain the higher-order curvature corrections.
Einstein-Gauss-Bonnet gravity is one of the natural modifications for Einstein gravity at short distances by including Gauss-Bonnet term which arises naturally from the low-energy limit of heterotic string theory \cite{Zwiebach1985,Witten1986,Gross1987}. Importantly, the presence of Gauss-Bonnet term does not lead to more than second derivatives of the metric in the corresponding field equations and thus the theory is ghost-free. Also, Gauss-Bonnet term only plays the role in spacetime with the dimension $d>4$. Whereas, for $d\leq4$, Gauss-Bonnet term is a topological invariant and thus it does not contribute to the field equations. Studies of holographic superconductors have been generalized in the framework of Einstein-Gauss-Bonnet gravity \cite{Gregory2009,Pavan2010,Wang2010,Barclay2010,Cai2011,Kanno2010,Kanno2011,Pan-Chen2011,Barclay2011,Li-Zhang2011,Wang2011,Jing-Chen2012,Cai-Zhang2013,Cui-Xue2013}.

In the string theory, the fundamental building blocks are the one-dimensional strings. Inspired by this,
the concept of string cloud was introduced as the one-dimensional analogous of a dust cloud. The black hole solution with the source of the string cloud was first found by Letelier \cite{Letelier1979}. Later, various black hole solutions in the presence of string cloud and their thermodynamic properties have been studied in the literature \cite{Herscovich2010,Baboolal2014}. 

Inspired by these ideals, in this paper we study holographic superconductors with involving simultaneously the higher-order curvature corrections and string cloud. In Sect. \ref{HDM}, we set up a gravitational dual model. The starting point is to construct a $d$-dimensional black brane solution sourced by a string cloud and negative cosmological constant in Einstein-Gauss-Bonnet gravity, which is the background geometry for studying holographic superconductors in the probe limit. The matter action and the corresponding equations of motion are also provided. In Sect. \ref{CTCD}, we obtain analytically the expression of the critical temperature in terms of the charge density, string cloud density parameter, Gauss-Bonnet coupling, and spacetime dimension. Furthermore, we obtain the expression of the condensation operator or order parameter near the critical temperature in Sect. \ref{CVCE}.

\section{\label{HDM} Holographic dual model}
In this section, we will build a holographic dual model which is used for the subsequent computation of critical phenomena. First let us introduce the action describing $d$-dimensional Einstein-Gauss-Bonnet gravity coupled to a electromagnetic field and scalar field surrounded by a string cloud in the AdS spacetime background as
\begin{equation}
S=\frac{1}{2}\int
d^dx\sqrt{-g}\left[R+\frac{(d-1)(d-2)}{l^2}+\alpha\mathcal{L}_{GB}+\mathcal{L}_m\right]+I_{\text{string-cloud}},\label{EGB-ED-adS}
\end{equation}
where $R$ is the scalar curvature of the spacetime, $l$ is the curvature radius of the AdS spacetime, and $\mathcal{L}_{GB}$ is the Gauss-Bonnet term given as
\begin{equation}
  \mathcal{L}_{GB}=R^2-4R_{\mu\nu}R^{\mu\nu}+R_{\mu\nu\rho\lambda}R^{\mu\nu\rho\lambda},
\end{equation}
and $\alpha$ is the Gauss-Bonnet coupling parameter. The matter term is given as
\begin{equation}
\mathcal{L}_m=-\frac{F_{\mu\nu}F^{\mu\nu}}{4}-|(\nabla_\mu-iqA_\mu)\psi|^2-m^2|\psi|^2,
\end{equation}
where $F_{\mu\nu}=\partial_\mu A_\nu-\partial_\nu A_\mu$ is the strength tensor of the electromagnetic field $A_\mu$, and $\psi$ refers to the scalar field of the charge $q$ and the mass $m$. $I_{\text{string-cloud}}$ is the action for the string cloud.

\subsection{Black brane solution with source of string cloud}
First let us review briefly the model of the string cloud \cite{Letelier1979}. For a classical relativistic string moving in the spacetime, its dynamics is described by the Nambu-Goto action $I_{\text{NG}}$ given by
\begin{equation}
I_{\text{NG}}=-T_p\int_{\Sigma}\sqrt{-\gamma}d\sigma^0d\sigma^1,
\end{equation}
where $T_p$ is the tension of the string and $(\sigma^0,\sigma^1)$ are the coordinates parameterizing the worldsheet $\Sigma$, $\gamma$ is the determinant of the induced metric $\gamma_{ab}=g_{\mu\nu}\frac{\partial x^\mu}{\partial\sigma^a}\frac{\partial x^\nu}{\partial\sigma^b}$. The worldsheet $\Sigma$ can be described by a bivector of the form
\begin{equation}
\Sigma^{\mu\nu}=\epsilon^{ab}\frac{\partial x^\mu}{\partial\sigma^a}\frac{\partial x^\nu}{\partial\sigma^b},
\end{equation}
where $\epsilon^{ab}$ is the two-dimensional Levi-Civita tensor, $\epsilon^{00}=\epsilon^{11}=0$ and $\epsilon^{01}=-\epsilon^{10}=1$. This bivector satisfies the identities, $\Sigma^{\mu[\rho}\Sigma^{\lambda\nu]}=0$ and $\nabla_\mu\Sigma^{\mu[\rho}\Sigma^{\lambda\nu]}=0$ (where the square brackets refer to antisymmetrization in the closed indices), which by the Frobenius's theorem it leads to a parameterized surface. Also, the bivector $\Sigma^{\mu\nu}$ satisfies another identity, $\Sigma^{\mu\rho}\Sigma_{\rho\lambda}\Sigma^{\lambda\nu}=\gamma\Sigma^{\mu\nu}$. In this description, the Lagrangian of the string is given by
\begin{equation}
\mathcal{L}_{\text{str}}=-T_p\left(-\frac{\Sigma^{\mu\nu}\Sigma_{\mu\nu}}{2}\right)^{1/2}.
\end{equation}
Since we can obtain the energy-momentum tensor for the string as $T^{\mu\nu}=-2\frac{\partial\mathcal{L}_{\text{str}}}{\partial g_{\mu\nu}}=T_p(-\gamma)^{-1/2}\Sigma^{\mu\sigma}{\Sigma_\sigma}^\nu$. 

In this work, we consider the string cloud with the energy-momentum tensor given by
\begin{equation}
T^{\mu\nu}=\rho\frac{\Sigma^{\mu\sigma}{\Sigma_\sigma}^\nu}{\sqrt{-\gamma}},
\end{equation}
where $\rho$ is the density of the string cloud. For the spherically symmetric and static string cloud, the ansatz of the bivector $\Sigma^{\mu\nu}$ is given as 
\begin{equation}
\Sigma^{\mu\nu}=B(r)\left({\delta}^\mu_t{\delta}^\nu_r-{\delta}^\nu_t{\delta}^\mu_r\right).
\end{equation} 
As a result, one can find the non-zero components of the energy-momentum tensor for the string cloud as
\begin{equation}
{T^t}_t={T^r}_r=-\rho|B(r)|.
\end{equation}
By using the relation $\partial_\mu\left(\sqrt{-g}\rho\Sigma^{\mu\nu}\right)=0$, we can obtain explicitly ${T^t}_t$ and ${T^r}_r$ as
\begin{equation}
{T^t}_t={T^r}_r=-\frac{a}{r^{d-2}},
\end{equation}
where $a$ is a positive real constant.

Varying the action (\ref{EGB-ED-adS}) with respect to the spacetime metric $g_{\mu\nu}$, we derive the equations of motion for $g_{\mu\nu}$ as
\begin{eqnarray}
{G^\mu}_\nu+\alpha{H^\mu}_\nu-\frac{(d-1)(d-2)}{2l^2}{\delta^\mu}_\nu&=&{T^\mu}_\nu(\text{matter})+{T^\mu}_\nu(\text{string-cloud}),\label{Eeq}
\end{eqnarray}
where
\begin{equation}
  {H^\mu}_\nu=2\left(R{R^\mu}_\nu-2R^{\mu\sigma}R_{\sigma\nu}-2R^{\sigma\rho}{R^\mu}_{\sigma\nu\rho}+{R^\mu}_{\rho\sigma\lambda}{R_\nu}^{\rho\sigma\lambda}\right)-\frac{1}{2}{\delta^\mu}_\nu\mathcal{L}_{GB}.
\end{equation}
In this work, we would like to investigate the role of the string cloud on holographic superconductors in Einstein-Gauss-Bonnet gravity. Thus, we shall consider the backreaction of the string cloud on the spacetime geometry. Whereas, the matter fields (the gauge field $A_\mu$ and the scalar field $\psi$) are considered in the probe limit which means that their backreaction or the matter term $\mathcal{L}_m$ on the spacetime geometry is ignored. In this sense, the spacetime geometry in Einstein-Gauss-Bonnet gravity is sourced by the negative cosmological constant and the string cloud. Whereas, the matter fields decouple to gravity. 

Now, we find a planar Schwarzschild-AdS black hole solution given by ansatz as
\begin{equation}
ds^2=-f(r)dt^2+\frac{dr^2}{f(r)}+r^2h_{ij}dx^idx^j,
\end{equation}
where $h_{ij}dx^idx^j=dx^2_1+dx^2_2+...+dx^2_{d-2}$ is the line element of the $(d-2)$-dimensional planar hypersurface. The $(t,t)$ component of Einstein's field equations leads to the equation for the function $f(r)$ as
\begin{equation}
\frac{d-2}{2r^4}\left\{r^2(d-3)f+r^3f'-\widetilde{\alpha}f\left[(d-5)f+2rf'\right]\right\}-\frac{(d-1)(d-2)}{2l^2}=-\frac{a}{r^{d-2}},
\end{equation}
where $\widetilde{\alpha}=\alpha(d-3)(d-4)$. By solving this equation, we can obtain the 
\begin{equation}
f(r)=\frac{r^2}{2\widetilde{\alpha}}\left(1-\sqrt{1-\frac{4\widetilde{\alpha}}{l^2}+\frac{4\widetilde{\alpha}m}{r^{d-1}}+\frac{8a\widetilde{\alpha}}{(d-2)r^{d-2}}}\right),
\end{equation}
where the reduced mass $m$ is related the total mass $M$ of the black hole as
\begin{equation}
M=\frac{(d-2)\omega_{d-2}}{16\pi}m,
\end{equation}
with $\omega_{d-2}$ to be the surface area of the unit $(d-2)$-sphere. The function $f(r)$ can be expressed in another form as
\begin{equation}
f(r)=\frac{r^2}{2\widetilde{\alpha}}\left(1-\sqrt{1-\frac{4\widetilde{\alpha}}{l^2}\left(1-\frac{r^{d-1}_+}{r^{d-1}}\right)+\frac{8a\widetilde{\alpha}}{(d-2)r^{d-2}}\left(1-\frac{r_+}{r}\right)}\right),
\end{equation}
where $r_+$ is the event horizon radius. The asymptotic behavior of the function $f(r)$, corresponding to $r\rightarrow\infty$, is given by
\begin{equation}
f(r)=\frac{r^2}{l^2_{\text{eff}}},
\end{equation}
where
\begin{equation}
l^2_{\text{eff}}=\frac{2\widetilde{\alpha}}{1-\sqrt{1-\frac{4\widetilde{\alpha}}{l^2}}},
\end{equation}
is the effective AdS radius. Note that, for the well-defined theory, the condition $\widetilde{\alpha}\leq l^2/4$ must be satisfied. The Hawking temperature of the black hole is given by
\begin{equation}
T_H=\frac{1}{4\pi}\left[\frac{(d-1)r_+}{l^2}-\frac{2a}{(d-2)r^{d-3}_+}\right],\label{bhtemp}
\end{equation}
which is interpreted as the temperature of the dual field theory.

\subsection{Basic setup for matter fields}
Varying the action (\ref{EGB-ED-adS}) with respect to the electromagnetic field $A_\mu$ and the scalar field $\phi$, we obtain
\begin{eqnarray}
\nabla^\nu F_{\mu\nu}+iq\left[\psi^*(\nabla_\mu-iqA_\mu)\psi-\psi(\nabla_\mu+iqA_\mu)\psi^*\right]&=&0,\nonumber\\
\left(\nabla_\mu-iqA_\mu\right)\left(\nabla^\mu-iqA^\mu\right)\psi-m^2\psi &=&0,\label{MFeq}
\end{eqnarray}
In order to solve these equations, we adopt spherically-symmetric and static ansatz for $A_\mu$ and $\psi$ as
\begin{equation}
A_\mu=\phi(r)\delta^t_\mu,\ \ \ \ \psi=\psi(r).
\end{equation}
By substituting this ansatz into Eq. (\ref{MFeq}), it leads to the equation for the functions $\phi(r)$ and $\psi(r)$ as
\begin{eqnarray}
\phi''(r)+\frac{d-2}{r}\phi'(r)-\frac{2q^2\phi(r)}{f(r)}\psi^2(r)&=&0,\label{r-phi-Eq}\\
 \psi''(r)+\left[\frac{f'(r)}{f(r)}+\frac{d-2}{r}\right]\psi'(r)+\left[\frac{q^2\phi^2(r)}{f^2(r)}-\frac{m^2}{f(r)}\right]\psi(r)&=&0,\label{r-psi-Eq}
\end{eqnarray}
where the prime is denoted the derivative with respect to the coordinate $r$. The regularity requirement for the fields $A_\mu$ and $\psi$ at the event horizon leads to the boundary conditions, $\phi(r_+)=0$ and $\psi(r_+)=\frac{f'(r_+)\psi'(r_+)}{m^2}$. Also, the behavior of $\phi(r)$ and $\psi(r)$ near the AdS boundary ($r\rightarrow\infty$) is given by
\begin{eqnarray}
\phi(r)&=&\mu-\frac{\rho}{r^{d-3}},\label{phi-asy-beh}\\
\psi(r)&=&\frac{\langle\mathcal{O}_-\rangle}{r^{\Delta_-}}+\frac{\langle\mathcal{O}_+\rangle}{r^{\Delta_+}},
\end{eqnarray}
where $\mu$ and $\rho$ are the chemical potential and the charge density,
respectively, and $\Delta_\pm=\left[(d-1)\pm\sqrt{(d-1)^2+4m^2l^2_{\text{eff}}}\right]/2$ are the conformal dimensions of $\langle\mathcal{O}_\pm\rangle$, respectively.\footnote{The mass of the scalar field satisfies the Breitenlohner-Freedman bound \cite{Breitenlohner1982}
\begin{equation}
m^2\geq-\frac{(d-1)^2}{4l^2_{\text{eff}}}.
\end{equation}} According to AdS/CFT correspondence, either $\langle\mathcal{O}_-\rangle$ or $\langle\mathcal{O}_+\rangle$ is interpreted as the source and the other is interpreted as the expectation value of the condensation operator in the boundary field theory.
In this work, we consider $\langle\mathcal{O}_+\rangle$ as  the expectation value of the condensation operator. Whereas $\langle\mathcal{O}_-\rangle$ is considered as the source which is set to be zero because the $\mathrm{U}(1)$ symmetry is broken spontaneously.

Interestingly, Eqs. (\ref{r-phi-Eq}) and (\ref{r-psi-Eq}) possess the following scaling symmetries
\begin{enumerate}
\item $\phi\rightarrow\lambda_1\phi,\ \ \psi\rightarrow\lambda_1\psi,\  \ q\rightarrow\lambda^{-1}_1q$.
\item $\phi\rightarrow\lambda_2\phi,\ \ \psi\rightarrow\lambda^{1/2}_2\psi,\ \ a\rightarrow\lambda_2a,\ \ \widetilde{\alpha}\rightarrow\lambda^{-1}_2\widetilde{\alpha},\ \ l\rightarrow\lambda^{-1/2}_2l,\ \ m\rightarrow\lambda^{1/2}_2m$,
\end{enumerate}
which can be used to set $q=l=1$.

With new coordinate $z=\frac{r_+}{r}$, Eqs. (\ref{r-phi-Eq}) and (\ref{r-psi-Eq}) are rewritten as 
\begin{eqnarray}
 \phi''(z)+\frac{4-d}{z}\phi'(z)-\frac{2 r^2_+\phi(z)}{z^4f(z)}\psi^2(z)&=&0,\label{z-phi-Eq}\\
 \psi''(z)+\left(\frac{f'(z)}{f(z)}+\frac{4-d}{z}\right)\psi'(z)+\frac{r^2_+}{z^4}\left(\frac{\phi^2(z)}{f^2(z)}-\frac{m^2}{f(z)}\right)\psi(z)&=&0,\label{z-psi-Eq}
\end{eqnarray}
where the prime is denoted the derivative with respect to the coordinate $z$, and
\begin{equation}
f(z)=\frac{r^2_+}{2\widetilde{\alpha}z^2}\left(1-\sqrt{1-4\widetilde{\alpha}\left(1-z^{d-1}\right)+\frac{8a\widetilde{\alpha}z^{d-2}}{(d-2)r^{d-2}_+}(1-z)}\right).
\end{equation}

\section{\label{CTCD} Critical temperature versus charge density}
In this section, we will obtain the expression for the critical temperature $T_c$ and investigate how it behaves under the change of the string cloud density parameter $a$, the spacetime dimension $d$ as well as the Gauss-Bonnet coupling $\alpha$. The analytical calculation is based on the Sturm-Liouville eigenvalue method. 

The condensation is zero at the critical temperature $T_c$, and thus the scalar field vanishes, $\psi=0$. Consequently, Eq. (\ref{z-phi-Eq}) yields the following equation
\begin{equation}
\phi''(z)+\frac{4-d}{z}\phi'(z)=0.
\end{equation}
The general solution of this equation is given as
\begin{equation}
\phi(z)=C_1+C_2z^{d-3}.\label{Tc-z-phi-sol1}
\end{equation}
By using the boundary condition at the horizon $\phi(1)=0$ and the asymtotic behavior at Eq. (\ref{phi-asy-beh}), we obtain
\begin{equation}
\phi(z)=\frac{\rho}{r^{d-3}_{+c}}\left(1-z^{d-3}\right),\label{Tc-z-phi-sol2}
\end{equation}
where $r_{+c}$ is denoted the event horizon radius of the black hole with the temperature $T_c$. 
The function $\phi(z)$ is expressed as follows
\begin{equation}
\phi(z)=\lambda r_{+c}\xi(z),\label{near-Ads-phi}
\end{equation}
where $\lambda=\frac{\rho}{r^{d-2}_{+c}}$ and $\xi(z)=1-z^{d-3}$. In the limit $T\rightarrow T_c$, we express $\psi(z)$ by
\begin{equation}
\psi(z)=\langle\mathcal{O}_+\rangle\frac{z^{\Delta_+}}{r^{\Delta_+}_+}F(z),\label{near-Ads-psi}
\end{equation}
where $F(z)$ is the trial function satisfying the boundary condition $F(0)=1$ and $F'(0)=0$. By substituting this expression of $\psi(z)$ and the solution of $\phi(z)$ at the temperature $T_c$ into Eq. (\ref{z-psi-Eq}),  
we obtain
\begin{equation}
F''(z)+p(z)F'(z)+q(z)F(z)+\lambda^2w(z)\xi^2(z)F(z)=0,
\end{equation}
where
\begin{eqnarray}
p(z)&=&\frac{4-d+2\Delta_+}{z}+\frac{g'(z)}{g(z)},\nonumber\\
q(z)&=&\frac{\Delta_+}{z}\left[\frac{3-d+\Delta_+}{z}+\frac{g'(z)}{g(z)}\right]-\frac{2\widetilde{\alpha}m^2}{z^4g(z)},\nonumber\\
w(z)&=&\frac{4\widetilde{\alpha}^2}{z^4g^2(z)},\nonumber\\
g(z)&=&\frac{1}{z^2}\left(1-\sqrt{1-4\widetilde{\alpha}\left(1-z^{d-1}\right)+\frac{8a\widetilde{\alpha}z^{d-2}(1-z)}{(d-2)r^{d-2}_+}}\right).
\end{eqnarray}
Interestingly, this equation can be written in the form of the Sturm-Liouville equation as
\begin{equation}
\left[T(z)F'(z)\right]'-Q(z)F(z)+\lambda^2P(z)F(z)=0, \label{SL-eq}
\end{equation}
where
\begin{eqnarray}
Q(z)&=&-T(z)q(z),\nonumber\\
P(z)&=&T(z)w(z)\xi^2(z),\nonumber\\
T(z)&=&e^{\int p(z)dz}=z^{2-d+2\Delta_+}\left[1-z^{d-1}-2b(1-z)z^{d-2}\right]\left\{
1-\left[2b(1-z)+z\right]z^{d-2}\widetilde{\alpha}\right.\nonumber\\
&&\left.-\left[2b(1-z)+z\right]\left(4bz^{d+1}-4bz^d-2z^{d+1}+3z^2\right)z^{d-4}\widetilde{\alpha}^2+\mathcal{O}\left(\widetilde{\alpha}^3\right)\right\},
\end{eqnarray}
where $b\equiv\frac{a}{(d-2)r^{d-2}_+}$. It is known from the Sturm-Liouville eigenvalue problem that the eigenvalues of Eq. (\ref{SL-eq}) are obtained by minimizing the following expression
\begin{equation}
\lambda^2=\frac{\int^1_0T(z)F'^2(z)dz+\int^1_0Q(z)F^2(z)dz}{\int^1_0P(z)F^2(z)dz},\label{sqlam-func}
\end{equation}
where the trial function $F(z)$ is chosen as $F(z)=1-\beta z^2$ \cite{Therrien2010}. With this result, from Eq. (\ref{bhtemp}), we derive the critical temperature $T_c$ which depends on the charge density $\rho$, string cloud density parameter $a$, Gauss-Bonnet coupling $\alpha$ and spacetime dimension $d$ as
\begin{equation}
T_c=\frac{1}{4\pi}\left[(d-1)\left(\frac{\rho}{\lambda_{\text{min}}}\right)^{\frac{1}{d-2}}-\frac{2a}{(d-2)}\left(\frac{\lambda_{\text{min}}}{\rho}\right)^{\frac{d-3}{d-2}}\right],\label{ctemp-chden}
\end{equation} 
where $\lambda_{\text{min}}$ is determined by minimizing the function $\lambda^2$ given in Eq. (\ref{sqlam-func}). This is one of the main results in this present work. Clearly, the effect of the string cloud on the critical temperature $T_c$ enters through both the eigenvalue $\lambda_{\text{min}}$ and the expression for $T_c$. Whereas, Gauss-Bonnet term affects indirectly the critical temperature $T_c$ entering through the eigenvalue $\lambda_{\text{min}}$. It should be noted that, as the critical temperature is low enough, which means that $T_c$ goes to the zero, from Eq. (\ref{bhtemp}) we have the following relation
\begin{eqnarray}
r^{d-2}_+\simeq\frac{2a}{(d-1)(d-2)},
\end{eqnarray}
by which the parameter $b$ in Eq. (39) becomes $b\simeq\frac{d-1}{2}$. This suggests that in this limit the change of the string cloud density parameter $a$ affects no longer $\lambda_{\text{min}}$. On the other hand, in this limit the change of the string cloud density parameter $a$ affects the critical temperature through that $a$ enters directly in the expression for $T_c$. With $a=0$ corresponding to the absence of the string cloud, we get the usual expression for the critical temperature of the holographic superconductors in Einstein-Gauss-Bonnet gravity as, $T_c=\frac{d-1}{4\pi}\left(\frac{\rho}{\lambda_{\text{min}}}\right)^{1/(d-2)}$. One can easily see that, from the expression (\ref{ctemp-chden}), the critical temperature $T_c$ is possibly negative. As a result, the presence of the string cloud leads to a fact that the critical temperature only exists in an allowed region of the parameter space, given by
\begin{equation}
\frac{\rho}{a\lambda_{\text{min}}}>\frac{2}{(d-1)(d-2)}.
\end{equation}
For example, with $d=5$, $\alpha=0.02$, $a=1.5$, $r_+=1$ and $\Delta_+=3$, the critical temperature only exists for the charge density satisfying, $\rho\gtrsim0.975$. In addition, as the string cloud density parameter $a$ increases, $\lambda^2_{\text{min}}$ should decrease and become negative above a critical value $a_{\text{crt}}$. For example, the critical value of  the string cloud density parameter is given by, $a_{\text{crt}}\approx5.55369$, with $d=5$, $\alpha=0.1$, $r_+=1$ and $\Delta_+=3$. Therefore, the presence of the string cloud with the sufficiently large density should prevent the existence of the critical temperature below which the superconducting phase will appear in the dual field theory.

In this way, in the presence of the string cloud, the critical temperature only exists if the string cloud density parameter $a$ is below a certain critical value. Note that, in order to form the superconducting state, it needs to have another condition which the chemical potential is above a critical value (for the charge density kept fixed). Thus, for the appearance of the superconducting state, the fact that the parameter $a$ is below a critical value can be considered as a necessary condition. Whereas, the fact that the chemical potential is above a critical value can be considered as a sufficient condition.

\begin{figure}[t]
 \centering
\begin{tabular}{cc}
\includegraphics[width=0.45 \textwidth]{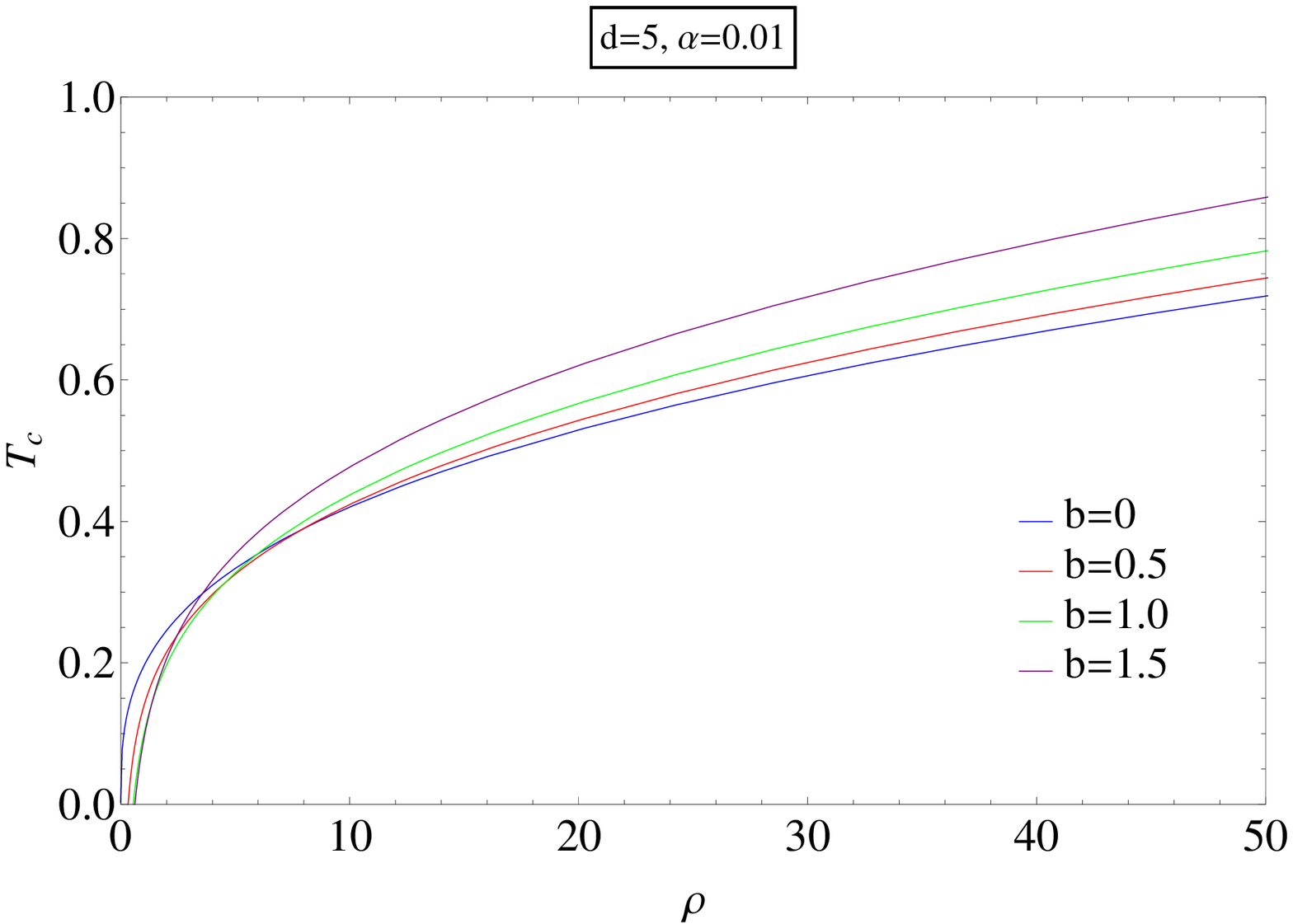}
\hspace*{0.05\textwidth}
\includegraphics[width=0.45 \textwidth]{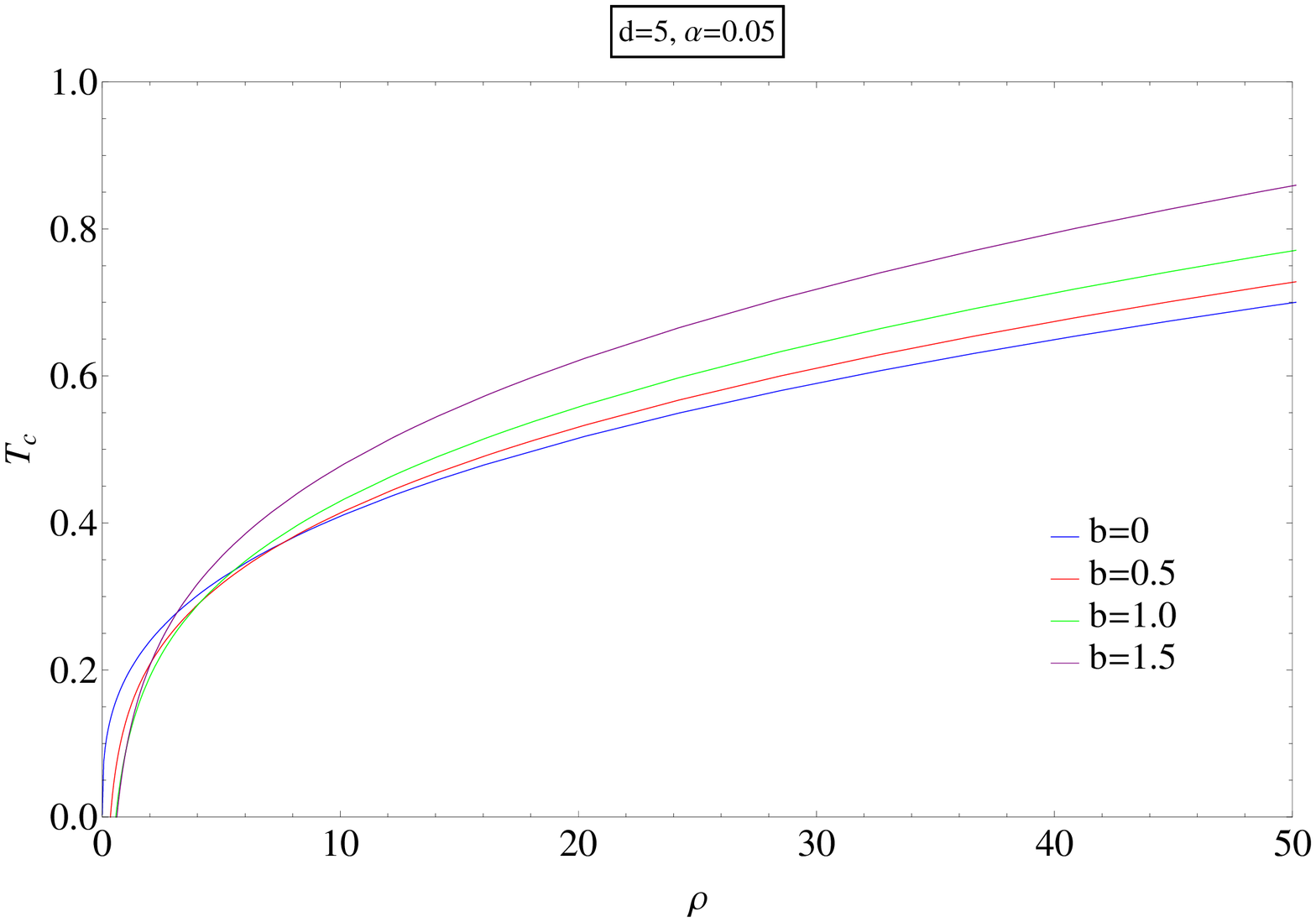}\\
\includegraphics[width=0.45 \textwidth]{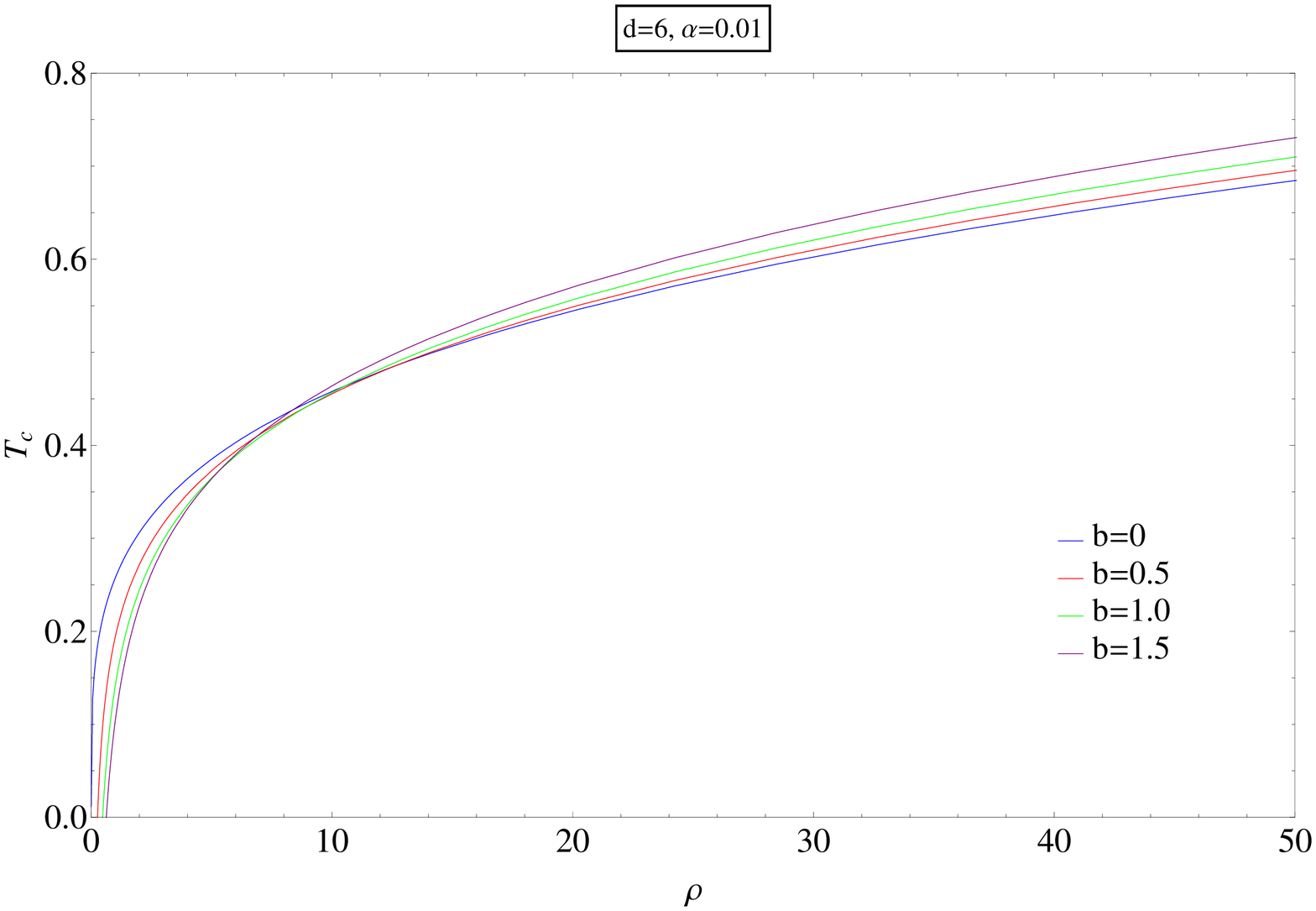}
\hspace*{0.05\textwidth}
\includegraphics[width=0.45 \textwidth]{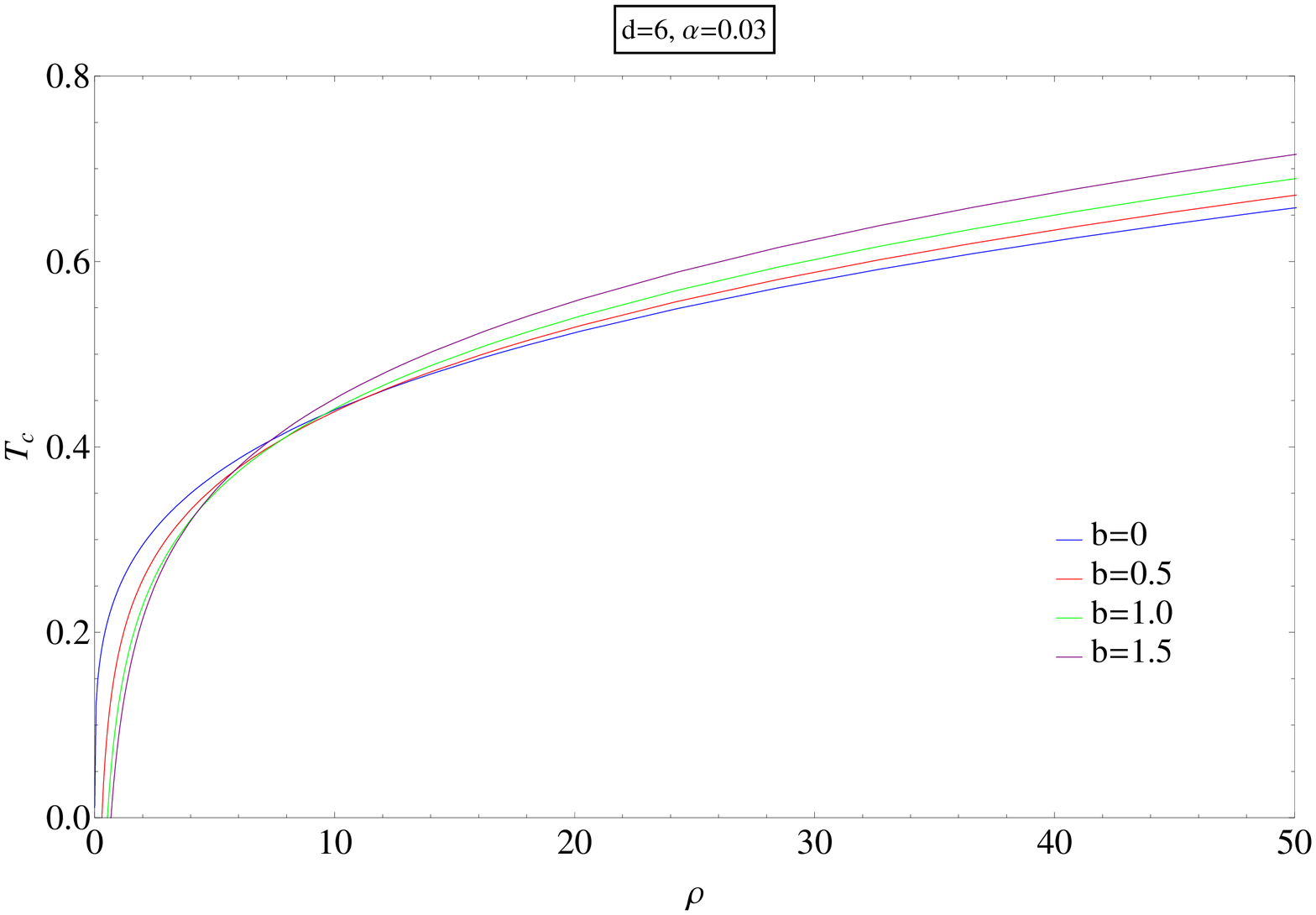}
\end{tabular}
  \caption{Plots of the critical temperature in terms of the charge density, at $m^2l^2_{\text{eff}}=-3$, where $b\equiv\frac{a}{(d-2)r^{d-2}_+}$.}\label{Tc-rho}
\end{figure}

\begin{figure}[t]
 \centering
\begin{tabular}{cc}
\includegraphics[width=0.6 \textwidth]{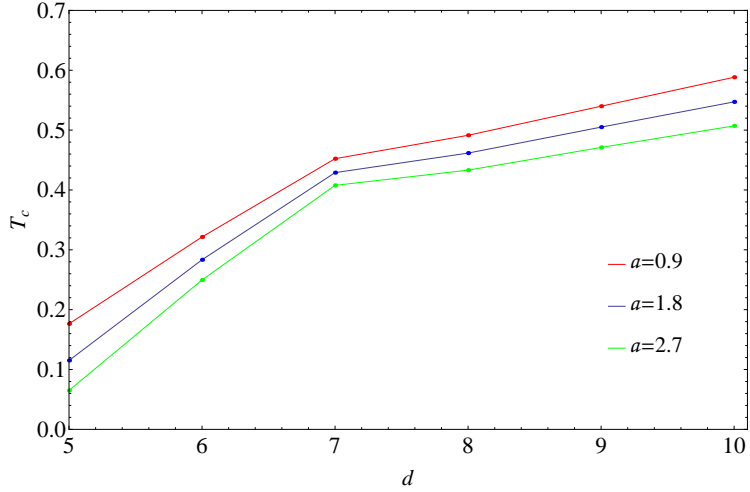}
\end{tabular}
  \caption{Plots of the critical temperature in terms of the spacetime dimension, at $\Delta_+=3$, $\rho=2$, $\alpha=0.005$ and $r_+=1$.}\label{Tc-dimension}
\end{figure}

In order to see more clearly, we plot the critical temperature $T_c$ as a function of the charge density $\rho$, for various values of $\alpha$, $a$ and $d$, in Fig. \ref{Tc-rho}.
We observe that, for the region of the sufficiently low charge density, the critical temperature decreases as increasing the string cloud density parameter $a$. This suggests that the condensation becomes harder in the presence of the string cloud. Whereas, for the region of the sufficiently high charge density, the critical temperature increases when the string cloud density parameter $a$ increases. And, since the presence of the string cloud makes the condensation getting easier. In addition, for studying the behavior of the critical temperature $T_c$ under the change of the spacetime dimension $d$, we plot $T_c$ as a function of $d$, for various values of $a$, in Fig. \ref{Tc-dimension}.
We can see from this figure that, in the presence of the string cloud, the critical temperature increases with the growth of the spacetime dimension. This is consistent with the previous works that showed superconductors of the high temperature can be achieved in the higher dimensional spacetime. Furthermore, we present a comparison between the results which are derived from the analytic and numerical calculations in Table. \ref{table-Tc-rho}, which shows the good agreement between two results. 
\begin{table}[!htp]
\centering
\begin{tabular}{|c|c|c|c|c|c|}
  \hline
  \multicolumn{3}{|c|}{$\alpha=0.01$} & \multicolumn{3}{|c|}{$\alpha=0.04$} \\
  \hline
  $b$ & $\text{Analytical}$ & $\text{Numerical}$ & $b$ & $\text{Analytical}$ & $\text{Numerical}$ \\
  \hline
  0 & $0.195\rho^{1/3}$ & $0.197\rho^{1/3}$ & 0 & $0.191\rho^{1/3}$ & $0.193\rho^{1/3}$  \\ \hline
 0.3 & $0.2\rho^{1/3}-0.04\rho^{-2/3}$ & $0.201\rho^{1/3}-0.04\rho^{-2/3}$ & 0.3 & $0.196\rho^{1/3}-0.042\rho^{-2/3}$ & $0.197\rho^{1/3}-0.041\rho^{-2/3}$  \\ \hline
  0.6 & $0.205\rho^{1/3}-0.076\rho^{-2/3}$ & $0.206\rho^{1/3}-0.076\rho^{-2/3}$ & 0.6 & $0.202\rho^{1/3}-0.079\rho^{-2/3}$ & $0.202\rho^{1/3}-0.078\rho^{-2/3}$  \\ \hline
  0.9 & $0.212\rho^{1/3}-0.108\rho^{-2/3}$ & $0.212\rho^{1/3}-0.107\rho^{-2/3}$ & 0.9 & $0.209\rho^{1/3}-0.11\rho^{-2/3}$ & $0.21\rho^{1/3}-0.11\rho^{-2/3}$ \\ \hline
  1.2 & $0.221\rho^{1/3}-0.132\rho^{-2/3}$ & $0.221\rho^{1/3}-0.132\rho^{-2/3}$ & 1.2 & $0.219\rho^{1/3}-0.134\rho^{-2/3}$ & $0.22\rho^{1/3}-0.133\rho^{-2/3}$ \\ \hline
  \end{tabular}
\caption{Analytical and numerical values of the critical temperature $T_c$, for the different values of $\alpha$ and $b\equiv\frac{a}{(d-2)r^{d-2}_+}$, at $d=5$ and $m^2l^2_{\text{eff}}=-3$ (or $\Delta_+=3$).}\label{table-Tc-rho}
\end{table}
Also, from this table, it is easily to see that the critical temperature decreases as increasing the Gauss-Bonnet coupling as indicated in the literature.

\section{\label{CVCE} Condensation value and critical exponent}
In this section, we would like to calculate the condensation value $\langle\mathcal{O}_+\rangle$ near the critical temperature $T_c$ and derive the corresponding critical exponent for Gauss-Bonnet holographic superconductors in the presence of the string cloud. In order to do this, we need to look at the behavior of the gauge field near $T_c$. It is important to note that near $T_c$ the condensation value $\langle\mathcal{O}_+\rangle$ is very small. This suggests that we can expand the function $\phi(z)$ in $\langle\mathcal{O}_+\rangle$ as
\begin{equation}
\phi(z)=\lambda r_+\left(1-z^{d-3}\right)+\frac{\langle\mathcal{O}_+\rangle^2}{r^{2\Delta_+}_+}r_+\chi(z)+\cdots,\label{phi-Exp}
\end{equation}
where the function $\chi(z)$ satisfies the boundary condition, $\chi(1)=\chi'(1)=0$. Then, substituting this expansion into Eq. (\ref{z-phi-Eq}), we obtain
\begin{equation}
\chi''(z)+\frac{4-d}{z}\chi'(z)-\frac{2\lambda r^2_+\left(1-z^{d-3}\right)}{f(z)}z^{-4+2\Delta_+}F^2(z)=0.\label{chi-Eq}
\end{equation}
This equation can be written as
\begin{equation}
\left[\frac{\chi'(z)}{z^{d-4}}\right]'=\frac{2\lambda r^2_+\left(1-z^{d-3}\right)}{f(z)}z^{-d+2\Delta_+}F^2(z).
\end{equation}
Using the condition $\chi'(1)=0$, we can integrate two sides of the above equation, which leads to
\begin{eqnarray}
\chi'(z)&=&z^{d-4}\int^{z}_1\frac{2\lambda r^2_+\left(1-\widetilde{z}^{d-3}\right)}{f(\widetilde{z})}\widetilde{z}^{-d+2\Delta_+}F^2(\widetilde{z})d\widetilde{z}.\label{Der-psi}
\end{eqnarray}

By identifying the asymptotic behavior of $\phi(z)$ from Eqs. (\ref{phi-asy-beh}) and (\ref{phi-Exp}), we get the following relation
\begin{equation}
\mu-\frac{\rho}{r^{d-3}_+}z^{d-3}=\lambda r_+\left(1-z^{d-3}\right)+\frac{\langle\mathcal{O}_+\rangle^2}{r^{2\Delta_+-1}_+}\left[\chi(0)+\chi'(0)z+\frac{\chi''(0)}{2}z^2+\cdots+\frac{\chi^{(d-3)}(0)}{(d-3)!}z^{d-3}+\cdots\right].
\end{equation}
Comparing the coefficients of the terms relating to $z^{d-3}$ in the right-hand and left-hand sides of the above equation, we find
\begin{equation}
\frac{\rho}{r^{d-2}_+}=\lambda-\frac{\langle\mathcal{O}_+\rangle^2}{r^{2\Delta_+}_+}\frac{\chi^{(d-3)}(0)}{(d-3)!}.\label{rel}
\end{equation}
It should also note that we can determine $\chi'(0)=\chi''(0)=\cdots=\chi^{(d-4)}(0)=0$ which is consistent to the expression of $\chi'(z)$ given in Eq. (\ref{Der-psi}). With the result (\ref{rel}), we find the expression for the condensation value $\langle\mathcal{O}_+\rangle$ as
\begin{eqnarray}
\langle\mathcal{O}_+\rangle=\frac{\Theta  T^{\Delta_+}}{\sqrt{d-2}}\sqrt{\left(\frac{T_c}{T}\right)^{d-2}-1}\approx\Theta  T^{\Delta_+}_c\sqrt{1-\frac{T}{T_c}},
\end{eqnarray}
where
\begin{equation}
\Theta=\left(\frac{4\pi}{d-1}\right)^{\Delta_+}\sqrt{-\frac{\lambda(d-2)!}{\chi^{(d-3)}(0)}}.
\end{equation}
This expression shows clearly that the critical exponent is $1/2$, which is consistent with the mean-field value and is independent on the string cloud density parameter, Gauss-Bonnet coupling and spacetime dimension. In this sense, the value $1/2$ of the critical exponent seems to be a universal constant. Also, the presence of the string cloud and Gauss-Bonnet term do not affect the expression form of the condensation operator, which is a universal property for holographic superconductors. On the other hand, the presence of the string cloud and Gauss-Bonnet term only affect indirectly $\langle\mathcal{O}_+\rangle$ through modifying the value of $T_c$ and $\Theta$.

In order to calculate $\Theta$, we use a fact that in the limit $z\rightarrow0$ Eq. (\ref{chi-Eq}) becomes
\begin{equation}
\chi''(0)=\frac{d-4}{z}\chi'(z)\Big|_{z\rightarrow0}.
\end{equation}
This leads to the following relation between the $(d-3)$-th and first-order derivatives of $\chi$ at $z=0$
\begin{equation}
\frac{\chi^{(d-3)}(0)}{(d-4)!}=\frac{\chi'(z)}{z^{d-4}}\Big|_{z\rightarrow0}.
\end{equation}
Using this relation with $\chi'(z)$ given at Eq. (\ref{Der-psi}), we obtain
\begin{equation}
\Theta=\left(\frac{4\pi}{d-1}\right)^{\Delta_+}\sqrt{\frac{(d-2)(d-3)}{\mathcal{A}}},
\end{equation}
where $\mathcal{A}$ is given by
\begin{eqnarray}
\mathcal{A}&=&\int^{1}_0\frac{2r^2_+\left(1-z^{d-3}\right)}{f(z)}z^{-d+2\Delta_+}F^2(z)dz,\nonumber\\
&=&4\widetilde{\alpha}\int^{1}_0\frac{z^{-d+2+2\Delta_+}(1-z^{d-3})(1-\beta z^2)^2}{1-\sqrt{1-4\widetilde{\alpha}\left(1-z^{d-1}\right)+8b\widetilde{\alpha}z^{d-2}(1-z)}}dz,\nonumber\\
&=&2\int^{1}_0\frac{z^{-d+2+2\Delta_+}(1-z^{d-3})(1-\beta z^2)^2}{1-z^{d-1}-2bz^{d-2}(1-z)}dz+\widetilde{\alpha}\left(\frac{2}{d-3-2\Delta_+}+\frac{1}{\Delta_+}\right.\nonumber\\
&&\left.+\frac{\beta(d-3)}{(5-d+2\Delta_+)(1+\Delta_+}\right)-2\widetilde{\alpha}^2\left[\frac{\beta}{d-5-2\Delta_+}+\frac{1}{3-d+2\Delta_+}\right.\nonumber\\
&&\left.+\frac{2b}{d-2+2\Delta_+}+\frac{1-2b}{d-1+2\Delta_+}-\frac{2\beta b}{d+2\Delta_+}
+\frac{(2b-1)\beta}{d+1+2\Delta_+}-\frac{2b}{1+2\Delta_+}\right.\nonumber\\
&&\left.-\frac{2\beta b}{3+2\Delta_+}-\frac{1}{2}\left(\frac{1}{\Delta_+}+\frac{1-2b-\beta}{1+\Delta_+}+\frac{(2b-1)\beta}{2+\Delta_+}\right)\right]+\mathcal{O}\left(\widetilde{\alpha}^3\right),
\label{Afun}
\end{eqnarray}
which can be, with the given parameters, integrated numerically. 

\begin{figure}[t]
 \centering
\begin{tabular}{cc}
\includegraphics[width=0.45 \textwidth]{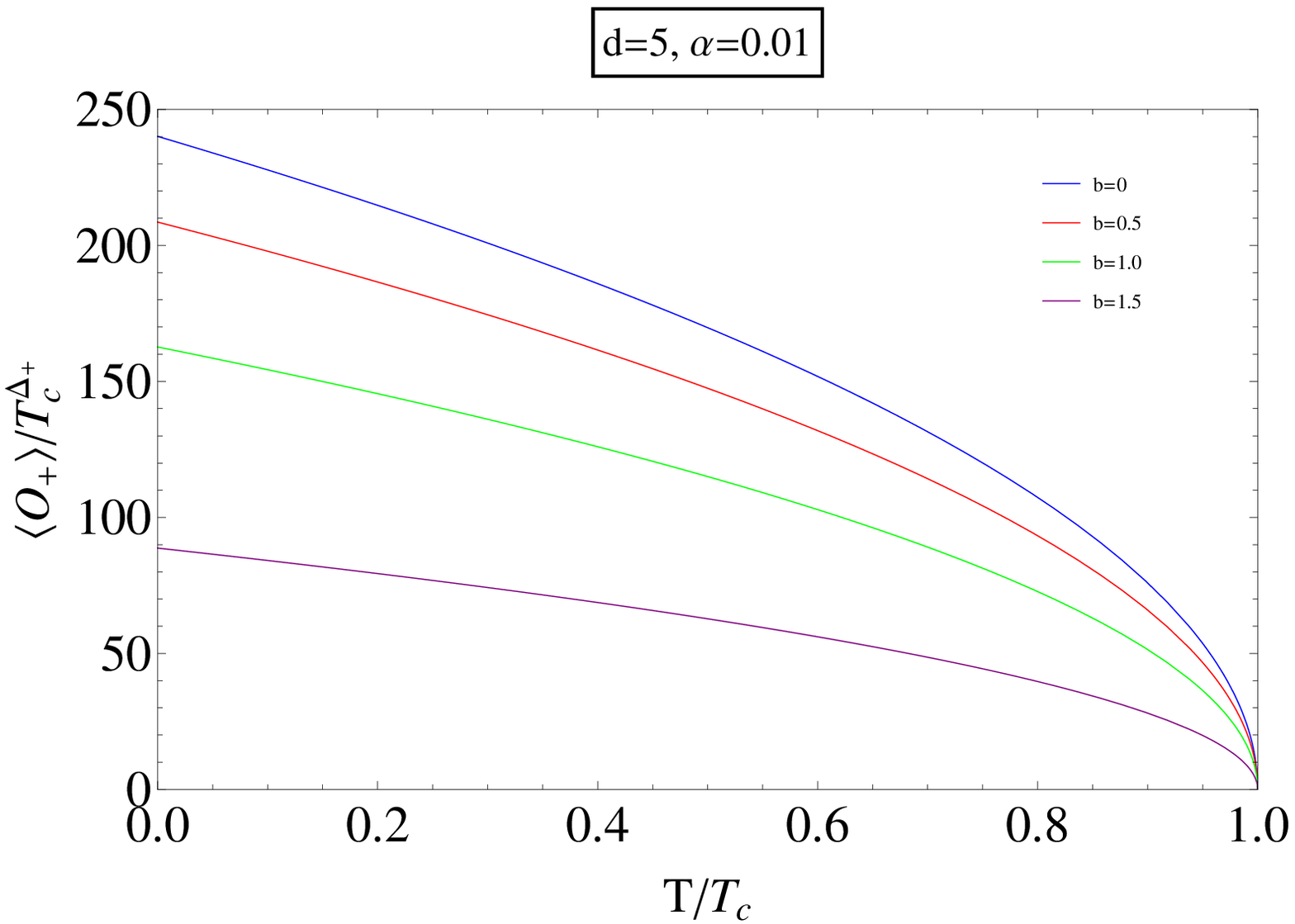}
\hspace*{0.05\textwidth}
\includegraphics[width=0.45 \textwidth]{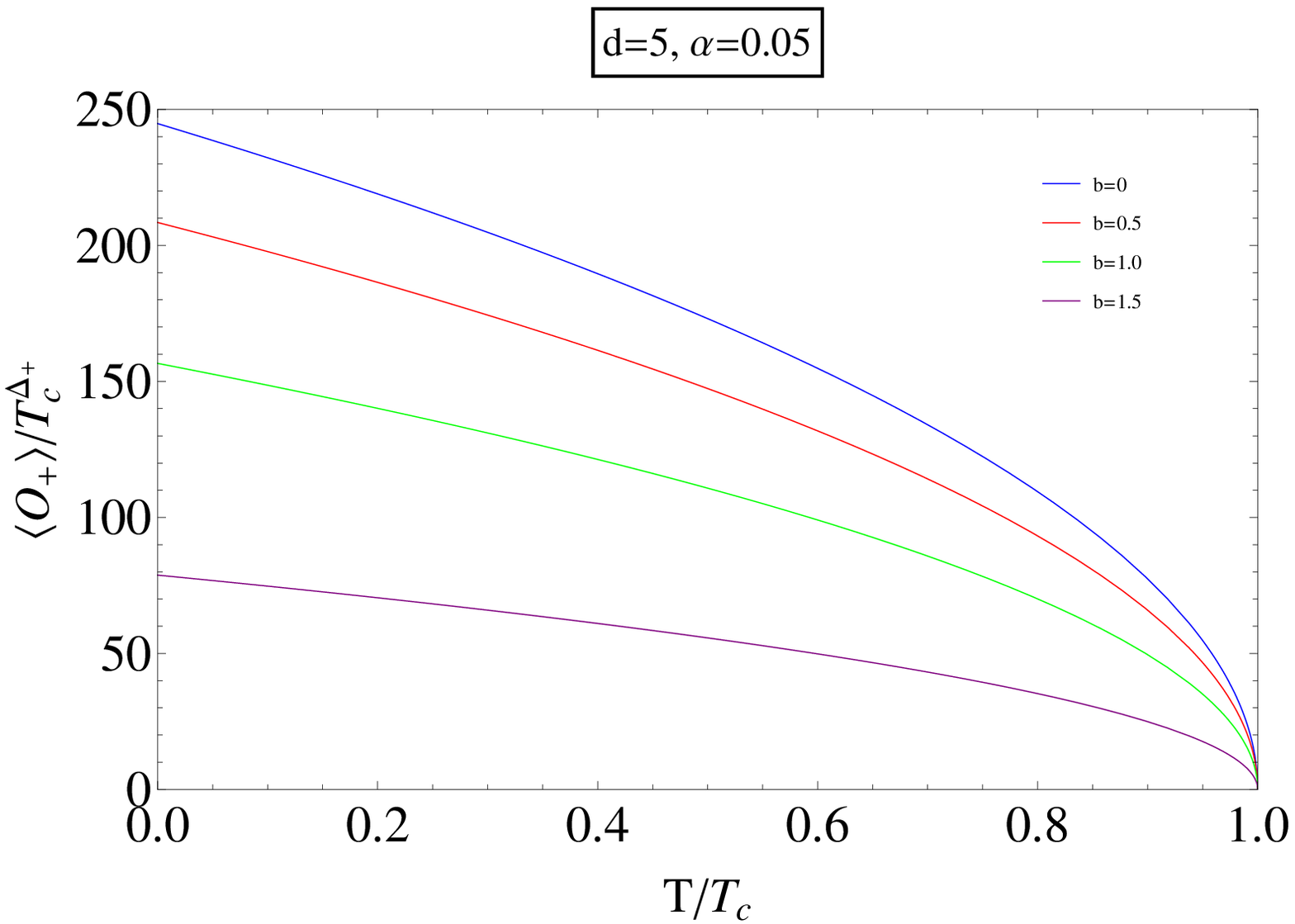}\\
\includegraphics[width=0.45 \textwidth]{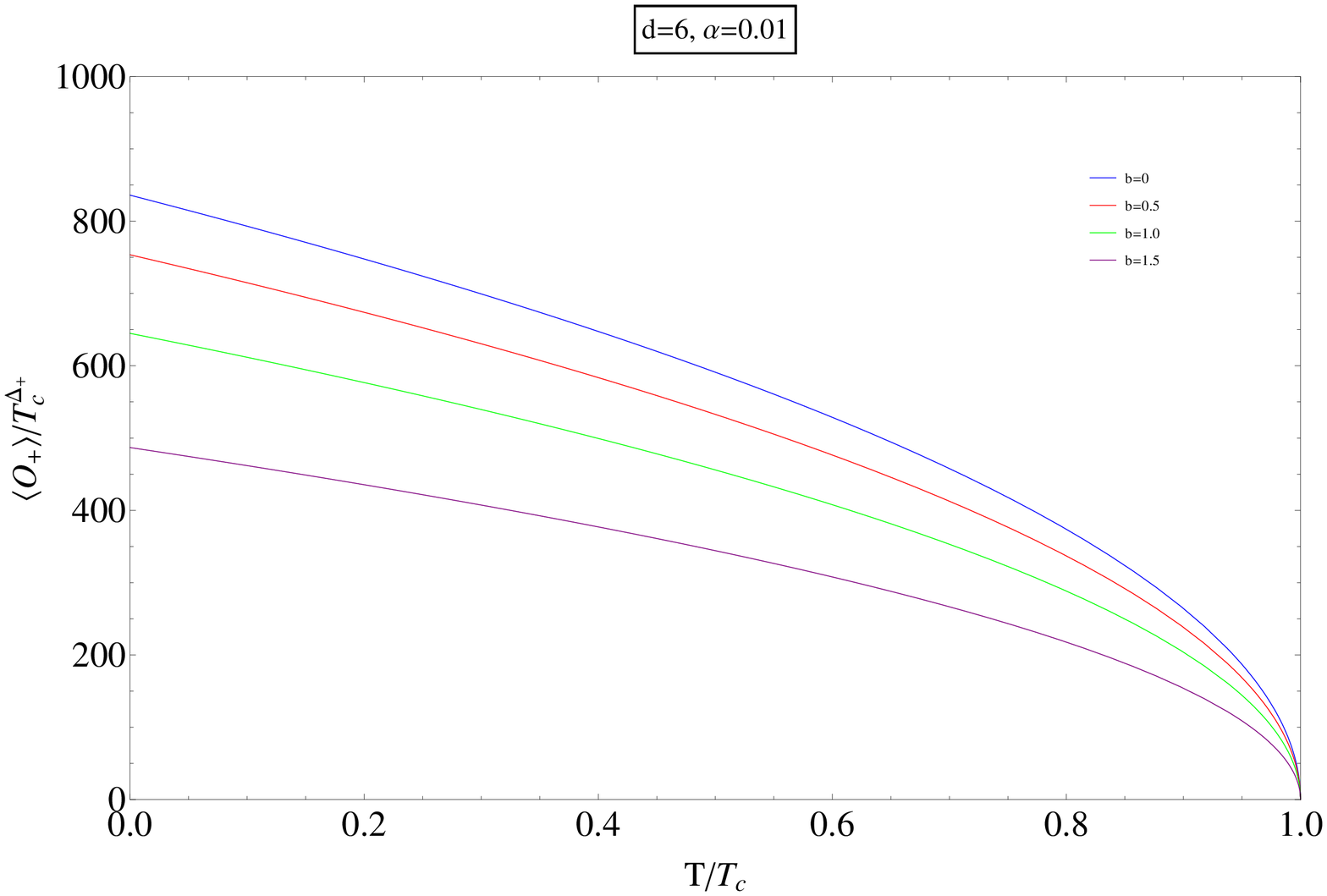}
\hspace*{0.05\textwidth}
\includegraphics[width=0.45 \textwidth]{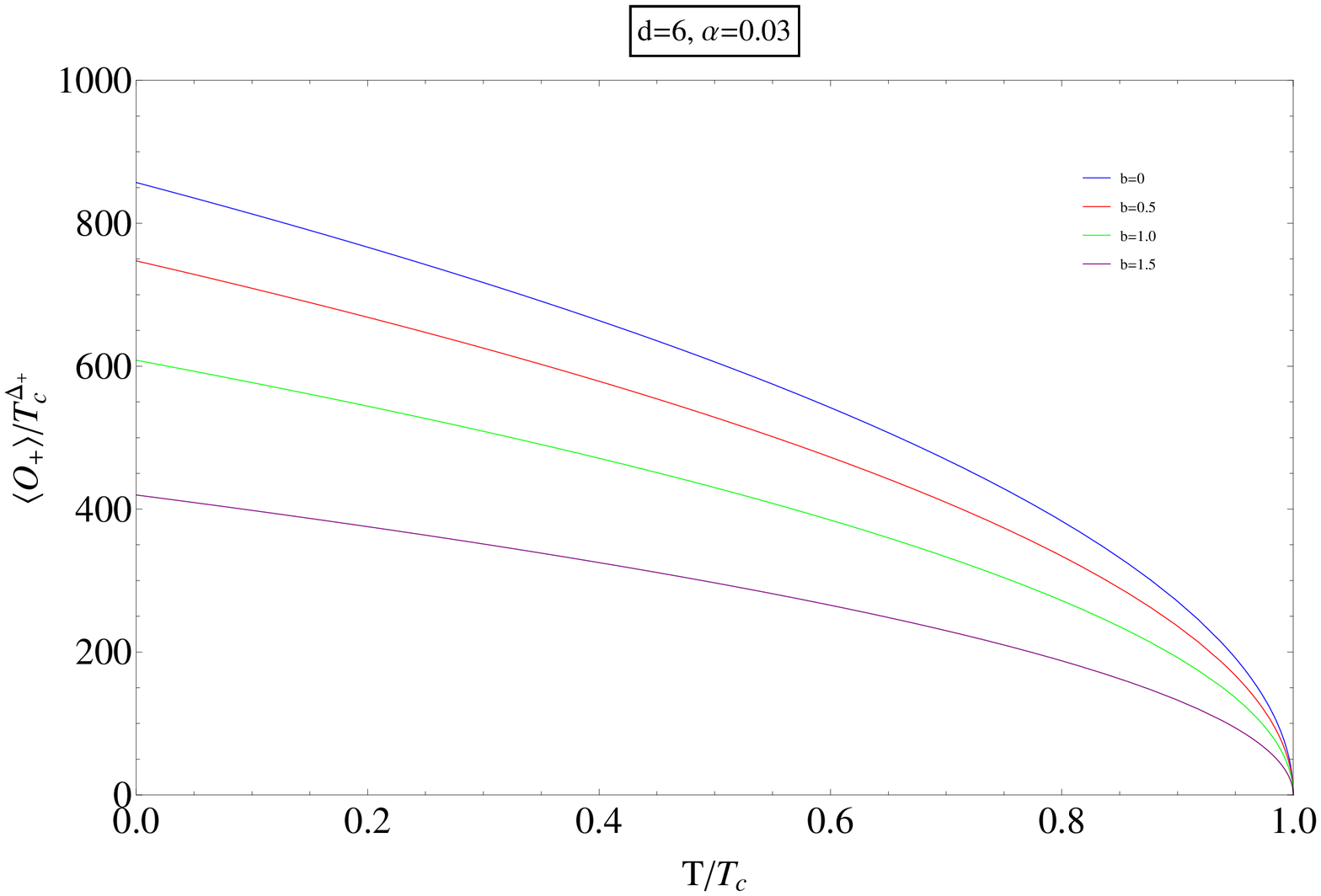}
\end{tabular}
  \caption{Plots of the dimensionless condensation operator in terms of $T/T_c$, at $m^2l^2_{\text{eff}}=-3$, where $b\equiv\frac{a}{(d-2)r^{d-2}_+}$.}\label{conds}
\end{figure}

With the expression (\ref{Afun}), we now are in position to perform the explicit calculation for the value of the dimensionless condensation operator $\langle\mathcal{O}_+\rangle/T^{\Delta_+}_c$. In Fig. \ref{conds}, we plot $\langle\mathcal{O}_+\rangle/T^{\Delta_+}_c$ as a function in terms of $T/T_c$ for various values of $\alpha$, $a$ and $d$. We observe that, for $T/T_c$ kept fixed, as the string cloud density parameter $a$ increases, the value of the dimensionless condensation operator becomes smaller. 

\section{\label{conclu} Conclusion}

In this paper, holographic superconductors have been investigated in the presence of a string cloud and Gauss-Bonnet term. In the probe limit at which the backreaction of the gauge field and scalar field on the background geometry is ignored all, we solve Einstein's field equation to obtain the planar AdS Gauss-Bonnet black hole solution sourced by the string cloud and negative cosmological constant. By adopting the Sturm-Liouville eigenvalue method, the critical temperature has been obtained as a function of the charge density, string cloud density parameter, Gauss-Bonnet coupling, and spacetime dimension. The presence of the string cloud modifies the usual form of the critical temperature (which is proportional to $\rho^{1/(d-2)}$). As a result, the existence of the critical temperature is only allowed in a certain region of the parameter space. In particular, the critical temperature, below which the condensation will be formed, only exists as the string cloud density parameter is smaller than a certain critical value. Also, it is indicated that when the string cloud density parameter increases, the critical temperature decreases in the region of the sufficiently low charge density, whereas it increases in the region of the sufficiently high charge density. In addition, the critical temperature growths with increasing the spacetime dimension, which suggests that the superconductors of the high temperature can be achieved in the higher spacetime dimension. It is also observed that increasing Gauss-Bonnet coupling always makes the critical temperature decreasing. Finally, the condensation operator and the critical exponent of holographic superconductors are computed at which the critical exponent is always equal to $1/2$ independently on the properties of the system.

\section*{Acknowledgments}
We would like to express sincere gratitude to the referees for their constructive comments and suggestions which have helped us to improve the quality of the paper.

\end{document}